\documentclass[preprint,tightenlines,showpacs,amsmath,amssymb]{revtex4}
\usepackage{graphicx}
\usepackage{dcolumn}
\usepackage{bm}
\usepackage{psfig}
\usepackage{rotating}
\newcommand{\dedx}{dE/dx}
\newcommand{\BR}{{\cal B}}
\newcommand{\eff}{\varepsilon}
\newcommand{\psip}{\psi^\prime}
\newcommand{\pspp}{\psi^{\prime\prime}}
\newcommand{\psp}{\psi^\prime}
\newcommand{\jpsi}{J/\psi}
\newcommand{\chicJ}{\chi_{cJ}}
\newcommand{\chicz}{\chi_{c0}}

\newcommand{\EE}{e^+e^-}
\newcommand{\MM}{\mu^+\mu^-}
\newcommand{\pip}{\pi^+}
\newcommand{\pim}{\pi^-}
\newcommand{\piz}{\pi^0}
\newcommand{\pp}{\pi^+\pi^-}
\newcommand{\rhopi}{\rho\pi}
\newcommand{\ppb}{p\overline{p}}
\newcommand{\RR}{R(2000)}
\newcommand{\ppjpsi}{\pi^+\pi^-\jpsi}
\newcommand{\kskl}{K^0_S K^0_L}
\newcommand{\ra}{\rightarrow}
\newcommand{\jpsito}{J/\psi \rightarrow }
\newcommand{\psipto}{\psi^\prime \rightarrow }
\newcommand{\pspto}{\psi^\prime \rightarrow }
\newcommand{\chicJto}{\chi_{cJ} \rightarrow }
\newcommand{\chiczto}{\chi_{c0} \rightarrow }

\newcommand{\bfg}{\begin{figure}}
\newcommand{\efg}{\end{figure}}
\newcommand{\bitm}{\begin{itemize}}
\newcommand{\eitm}{\end{itemize}}
\newcommand{\bnum}{\begin{enumerate}}
\newcommand{\enum}{\end{enumerate}}
\newcommand{\btbl}{\begin{table}}
\newcommand{\etbl}{\end{table}}
\newcommand{\btbu}{\begin{tabular}}
\newcommand{\etbu}{\end{tabular}}

\newcommand{\KK}{K^+K^-}

\newcommand{\beq}{\begin{equation}}
\newcommand{\edq}{\end{equation}}
\newcommand{\g}{\gamma}

\begin{document}
\normalsize
\parskip=5pt plus 1pt minus 1pt

\title{\boldmath Observation of $\ppb\piz$ and $\ppb\eta$ in $\psp$ decays}
\author{
M.~Ablikim$^{1}$,      J.~Z.~Bai$^{1}$,       Y.~Ban$^{11}$,
J.~G.~Bian$^{1}$,      X.~Cai$^{1}$,          J.~F.~Chang$^{1}$,
H.~F.~Chen$^{17}$,     H.~S.~Chen$^{1}$,      H.~X.~Chen$^{1}$,
J.~C.~Chen$^{1}$
% ...
\\(BES Collaboration)\\
$^{1}$ Institute of High Energy Physics, Beijing 100049,
People's Republic of China\\
% ...
$^{11}$ Peking University, Beijing 100871, People's Republic of China\\
% ...
$^{17}$ University of Science and Technology of China, Hefei
230026, People's Republic of China\\
% ...
}
\date{\today}

\begin{abstract}

The processes $\psipto \ppb \piz$ and $\psipto \ppb \eta$ are
studied using a sample of $14 \times 10^6$ $\psip$ decays
collected with the Beijing Spectrometer at the Beijing
Electron-Positron Collider. The branching fraction of $\psipto
\ppb \piz$ is measured with improved precision as $(13.2\pm 1.0\pm
1.5)\times 10^{-5}$, and $\psipto \ppb \eta$ is observed for the
first time with a branching fraction of $(5.8\pm 1.1\pm 0.7)\times
10^{-5}$, where the first errors are statistical and the second
ones are systematic.

\end{abstract}

\pacs{13.25.Gv, 12.38.Qk, 14.20.Gk, 14.40.Cs}

\maketitle

\section{Introduction}

There are some long-standing puzzles in the decays of vector
charmonia, in particular the ``$\rhopi$ puzzle'' between $\psp$
and $\jpsi$ decays and the possible large charmless decay
branching fraction of the $\pspp$. Following the suggestion in
Ref.~\cite{rosnersd} that the small $\pspto \rhopi$ branching
fraction is due to the cancellation of the $S$- and $D$-wave
matrix elements in $\psp$ decays, it was pointed out that all
$\psp$ decay channels should be affected by the same $S$- and
$D$-wave mixing scheme, and thus, in general, the ratios between
the branching fractions of $\psp$ and $\jpsi$ decay into the same
final states may have values different from the ``12\% rule'',
expected for pure $1S$ and $2S$ states~\cite{wymcharmless}. This
scenario also predicts $\pspp$ decay branching fractions since
$\pspp$ would also be a mixture of $S$- and $D$-wave charmonia,
and it was suggested that many $\jpsi$ and $\psp$, as well as
$\pspp$, decays should be measured to test this. For the channels
that have been measured, $\psp$ decays are found to be either
suppressed (like vector-pseudoscalar, vector-tensor), or enhanced
(like $\kskl$), or obey the 12\% rule (like baryon-antibaryon
pairs). In this paper, we analyze three-body decays of $\psp$ into
a $\ppb$ pair and a $\piz$ or $\eta$.

The $\jpsi$ and $\psp$ decays into $\ppb\piz$ and $\ppb\eta$ are
expected to be dominated by two-body decays involving excited
nucleon states. These states play an important role in the
understanding of nonperturbative
QCD~\cite{role1,role2,role3,role4}. However, our knowledge on
$N^*$ resonances, based on $\pi N$ and $\gamma N$
experiments~\cite{pdg}, is still very limited and imprecise.
Studies of $N^*$ resonances have also been performed using $\jpsi$
events collected at the Beijing Electron-Positron
Collider~(BEPC)~\cite{zounstar,jpsippeta,pnbarpi}, which provides
a new method for probing $N^*$ physics. Recently, based on 58
million $\jpsi$ events collected by BEijing Spectrometer (BESII),
a new $N^*$ peak with a mass at around 2065~MeV/$c^2$ was
observed~\cite{pnbarpi}. This may be one of the ``missing $N^*$
states'' around 2~GeV/$c^2$ that have been predicted by the quark
model in many of its forms~\cite{role2,model1,model2}.  However,
due to its large mass, the production of this $N^*(2065)$ in
$\jpsi$ decays is rather limited in phase space, and a similar
search for it in $\psp$ decays, which has a larger phase space
available may be helpful, although the production rate may be
small due to the large decay rates for $\psp$ into final states
with charmonium.

In a recent paper~\cite{zhangzx}, it is predicted that $\ppb$ in
$\pspto \ppb \piz$ may form iso-vector bound states near
threshold. These states can also be searched for.

Experimentally, $\psipto \ppb \piz$ was studied by Mark-II in
1983, with only 9 events observed~\cite{markii}, and the branching
ratio was found to be $(1.4\pm 0.5) \times 10^{-4}$. $\psipto \ppb
\eta$ has not been observed before.

\section{Detector and data samples}

The data used in this analysis were taken with the BESII detector
at the BEPC storage ring at a center-of-mass energy corresponding
to $M_{\psip}$. The data sample corresponds to a total of ($14.0
\pm 0.6)\times 10^6$ $\psip$ decays, as determined from inclusive
hadronic events~\cite{pspscan}.

BES is a conventional solenoidal magnet detector that is described
in detail in Refs.~\cite{bes,bes2}. A 12-layer vertex chamber
(VTC) surrounding the beam pipe provides trigger information. A
forty-layer main drift chamber (MDC), located radially outside the
VTC, provides trajectory and energy loss ($dE/dx$) information for
charged tracks over $85\%$ of the total solid angle.  The momentum
resolution is $\sigma _p/p = 0.017 \sqrt{1+p^2}$ ($p$ in
$\hbox{GeV}/c$), and the $dE/dx$ resolution for hadron tracks is
$\sim 8\%$. An array of 48 scintillation counters surrounding the
MDC measures the time-of-flight (TOF) of charged tracks with a
resolution of $\sim 200$ ps for hadrons.  Radially outside the TOF
system is a 12 radiation length, lead-gas barrel shower counter
(BSC).  This measures the energies of electrons and photons over
$\sim 80\%$ of the total solid angle with an energy resolution of
$\sigma_E/E=22\%/\sqrt{E}$ ($E$ in GeV). Outside of the solenoidal
coil, which provides a 0.4~Tesla magnetic field over the tracking
volume, is an iron flux return that is instrumented with three
double layers of counters that identify muons of momentum greater
than 0.5~GeV/$c$.

Monte Carlo (MC) simulation is used for the determination of the
invariant mass resolution and detection efficiency, as well as the
study of background. The simulation of the BESII detector is
Geant3 based, where the interactions of particles with the
detector material are simulated. Reasonable agreement between data
and Monte Carlo simulation is observed~\cite{simbes} in various
channels tested including $\EE \ra (\g)\EE$, $\EE\ra (\g)\MM$,
$\jpsito \ppb$ and $\psipto \ppjpsi$, $\jpsito \ell^+\ell^-$
$(\ell=e,\mu)$.

The signal channels $\psipto \ppb \piz$, $\piz \to 2\gamma$ and
$\psipto \ppb \eta$, $\eta \to 2\gamma$ are generated with a phase
space generator, giving similar $p\pi$, $\overline{p}\pi$,
$p\eta$, and $\overline{p}\eta$ mass distributions to those
observed in data.  For $\piz$ and $\eta$ channels, 100\,000 events
each are simulated. To study possible background in our analysis,
20\,000 $\pspto \piz \piz \jpsi$, $\jpsito \ppb$, 20\,000 $\psipto
\piz \piz \ppb$, and 30\,000 $\pspto \g \chicJ$, $\chicJto \g
\jpsi$, $\jpsito \ppb$ ($J=0,1,2$) events are generated.

\section{Event Selection}

The final states in which we are interested contain two photons
and two charged tracks. The number of charged tracks is required
to be two with net charge zero.  Each track should have good
quality in track fitting and satisfy $|\cos\theta|<0.8$, where
$\theta$ is the polar angle of the track measured by the MDC.

A neutral cluster in the BSC is considered to be a photon
candidate when the angle between the nearest charged track and the
cluster in the $xy$ plane is greater than $15^{\circ}$, the first
hit appears in the first five layers of the BSC (about six
radiation lengths of material), and the angle between the cluster
development direction in the BSC and the photon emission direction
in $xy$ plane is less than $37^{\circ}$. Two or three photon
candidates are allowed in an event, but the two with the largest
energies are selected as $\pi^0$ or $\eta$ decay candidates, and
both of them are required to have energy greater than 60~MeV.

A likelihood method is used for discriminating pion, kaon, proton,
and antiproton tracks. For every charged track, we define an
estimator as $E^{i}=P^{i}/\sum_{i} P^{i}$, where $P^{i}$ is the
probability under the hypothesis of being type $i$, $i=3,4,5$ for
$\pi, K$ and p or $\overline{p}$ hypotheses, respectively, and
$P^{i}=\prod_{j} P_{j}^{i}(x_{j})$. Here $P_{j}^{i}$ is the
probability density for the hypothesis of type $i$, associated to
the discriminating variable $x_{j}$. Discriminating variables used
for each charged track are time of flight in the TOF (TOF-T) and
the pulse height in the MDC ($\dedx$). By definition, pion, kaon,
proton, and antiproton tracks have corresponding $E^{i}$ values
near one. In this analysis, both tracks are required to have
$E^{5}>0.6$.

A four-constraint (four momentum conservation) kinematic fit is
made with the two charged tracks and two photon candidates; the
confidence level of the $\chi^2$ fit is required to be greater
than 1\%. A similar fit assuming the two charged tracks are $\KK$
is also performed, and the $\chi^2$ of $\psipto \gamma\gamma\ppb$
should be smaller than that of $\psipto \gamma\gamma\KK$.

\begin{figure*}[htbp]
\centerline{\hbox{
\psfig{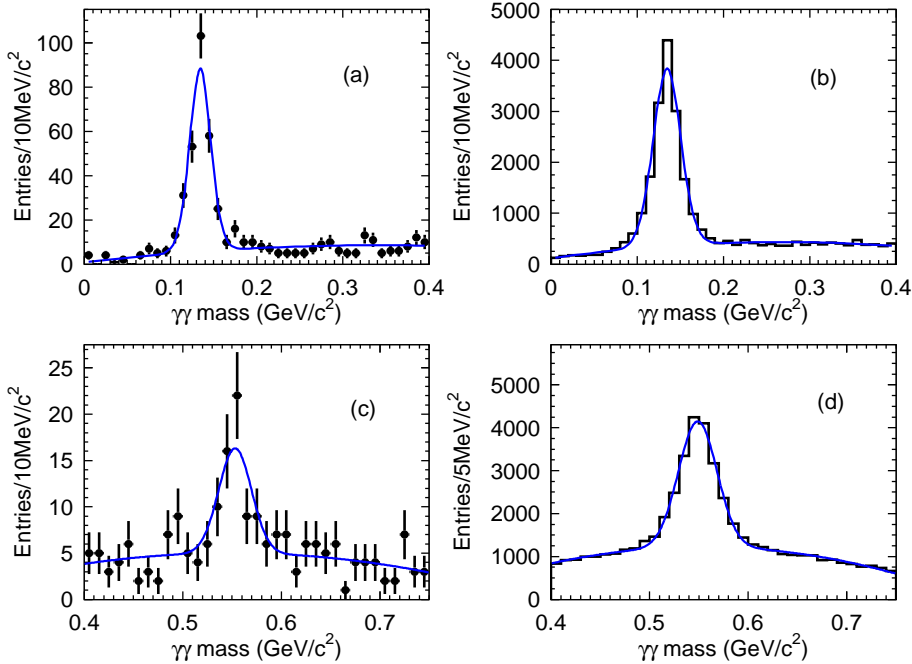}}}
\caption{Invariant mass distributions of $\gamma\gamma$ from the
selected $\pspto \g\g \ppb$ candidate events in data and in Monte
Carlo simulation described in Section V: (a) $\ppb\piz$ data, (b)
$\ppb\piz$ Monte Carlo simulation, (c) $\ppb\eta$ data, and (d)
$\ppb\eta$ Monte Carlo simulation. The curves show the best fit to
the distributions.} \label{mgg}
\end{figure*}

The scatter plot of the $\ppb$ invariant mass versus that of the
two photon candidates for events satisfying the above selection
criteria is shown in Fig.~\ref{jpsi}(a). The two bands with
$m_{\gamma\gamma}$ values near 0.135~GeV/$c^2$ and 0.547~GeV/$c^2$
are $\psipto \ppb \piz$, $\piz \to 2\gamma$ and $\psipto \ppb
\eta$, $\eta \to 2\gamma$ candidates, respectively. The band
corresponding to $\ppb$ mass around 3.1~GeV/$c^2$ is from $\pspto
\g \chicJ$, $\chicJto \g \jpsi$, $\jpsito \ppb$ ($J=0,1,2$), and
$\pspto \eta \jpsi$, $\jpsito \ppb$. The broadly distributed
background in the figure is due mainly to $\psipto \piz \piz
\jpsi$, $\jpsito \ppb$. Fig.~\ref{jpsi}(b) shows the corresponding
Monte Carlo distribution of these background channels, using
branching fractions measured by previous experiments~\cite{pdg}.
To remove these background events, we require $|m_{\ppb}-3.097|>$
1.5~$\sigma$ for $\g\g$ invariant mass ($m_{\g\g}$) smaller than
0.4~GeV/$c^2$, and $m_{\ppb}<3.2-0.3 m_{\g\g}$ for $m_{\g\g}$
larger than 0.4~GeV/$c^2$, where $\sigma\approx 0.011$~GeV/$c^2$
is the $\ppb$ invariant mass resolution as estimated from Monte
Carlo simulation of $\pspto \g\chicJ$. After the above selection,
the $\g\g$ invariant mass distributions are shown in
Figs.~\ref{mgg}(a) and (c), where clear $\piz$ and $\eta$ signals
are observed.

\begin{figure*}[htbp]
\centerline{\hbox{ \psfig{file=jpsi_ycz_fig1.epsi,width=12.0cm}}}
\caption{Scatter plots of $\ppb$ invariant mass versus $\g\g$
invariant mass before removing $\jpsi$ background. (a) is from
data and (b) is from Monte Carlo simulated $\pspto \g \chicJ$,
$\chicJto \g \jpsi$, $\jpsito \ppb$ ($J=0,1,2$), $\pspto \eta
\jpsi$, $\jpsito \ppb$, and $\psipto \piz \piz \jpsi$, $\jpsito
\ppb$ events. The lines show the selection criterion described in
the text.} \label{jpsi}
\end{figure*}

The same analysis is performed on a MC sample of 14~M inclusive
$\psip$ decays generated with Lundcharm~\cite{lundcharm}. It is
found that the remaining backgrounds are mainly from $\psipto \piz
\piz \ppb$, many via resonances such as $f_0$ and $f_2$. Some
other decay channels of $\psip$ with three photons such as
$\psipto \gamma \chicz, \chiczto \overline{p} \Delta^+, \Delta^+
\rightarrow \piz p$ are also observed. Since there are no
branching fractions  available for normalization of these
channels, we do not try to simulate all possible background
channels. Instead, in our fit to the $\g\g$ mass distributions, we
approximate the background shape by a smooth curve as predicted by
the inclusive MC sample.

\section{\boldmath Fits to the $\g\g$ mass distributions}

The $\g\g$ invariant mass distribution of candidates is not
described well by a simple Gaussian, but by the sum of multiple
Gaussians with different standard deviations, which depend on the
momentum of the $\piz$ or $\eta$. The analysis of the $\piz$
signal is done using five different momentum bins, which are fit
with multiple Gaussians for the signal and a second order
polynomial for the background. Summing up all the fits yields the
curve in Fig.~\ref{mgg}(a), and the total number of $\pspto \ppb
\piz$ events is found to be $256\pm 18$, with the error determined
from the fit.

The number of $\pspto \ppb \eta$ events is even more limited than
$\pspto \ppb \piz$, and we do not do the fit in $\eta$ momentum
bins. Instead the $\g\g$ invariant mass spectrum is fit with a
single Gaussian for the signal plus a second-order polynomial for
the background. In the fit, the mass resolution is fixed to
14.3~MeV/$c^2$, which is determined from Monte Carlo simulation
but calibrated using the $\piz$ signal in the $\ppb\piz$ fit. The
fit is shown in Fig.~\ref{mgg}(c), and the number of events is
found to be $44.8^{+8.7}_{-8.4}$. The $\eta$ mass from the fit,
$(552.4\pm 3.2)$~MeV/$c^2$, agrees well with the world
average~\cite{pdg}. The statistical significance of the $\ppb\eta$
signal is estimated to be 6.1$\sigma$ by comparing the likelihoods
with and without the signal in the fit.

\section{Resonance analysis and efficiency}\label{effi}

In order to determine the selection efficiency, it is necessary to
know the intermediate states in the decays for Monte Carlo
simulation. Figs.~\ref{dalitz}(a) and (c) are the Dalitz plots for
$\pspto \ppb \piz$ and $\pspto \ppb \eta$ after requiring the
$\g\g$ invariant mass to be consistent with a $\piz$
(0.11~GeV/$c^2<m_{\g\g}<0.16$~GeV/$c^2$) or $\eta$
(0.53~GeV/$c^2<m_{\g\g}<0.57$~GeV/$c^2$). In these figures, the
requirement on the $\ppb$ mass is removed to see the effect of the
backgrounds remaining from $\pspto \eta \jpsi$, $\pspto \piz\piz
\jpsi$, and $\pspto \gamma \chicJ$ in the lower left of the Dalitz
plots. These two figures differ significantly from phase space.
However, the data samples here are too small to perform a partial
wave analysis.

\begin{figure*}[htbp]
\centerline{\hbox{
\psfig{file=dalitz_ycz_fig3.epsi,width=14.0cm}}} \caption{Dalitz
plots for $\pspto \ppb \piz$ and $\pspto \ppb\eta$. (a) and (b)
are for $\pspto \ppb \piz$ data and the mixed Monte Carlo sample,
respectively, and (c) and (d) are for $\pspto \ppb \eta$ data and
the mixed Monte Carlo sample, respectively.} \label{dalitz}
\end{figure*}

In the MC simulation, $N^*(1535)\overline{p}+c.c.$ and $\RR\piz$ were
used in the $\ppb\piz$ mode and $N^*(1535)\overline{p}+c.c.$ and
$\RR\eta$ in $\ppb\eta$ mode, where $\RR$ is a state representing the
accumulation of the events near $\ppb$ mass threshold with
$$p\overline{N}^*(1535):\overline{p}N^*(1535):\piz \RR=2:2:1$$
for the $\ppb \piz$ mode, and
$$p\overline{N}^*(1535):\overline{p}N^*(1535):\eta \RR=5:5:3$$
for the $\ppb \eta$ mode, and $N^*$, $\piz$, and $\eta$ only decay
into desired final states. The MC simulations of the Dalitz plots
are shown in Figs.~\ref{dalitz}(b) and (d). The agreement between
data and MC simulation is reasonable.

\begin{figure*}[htbp]
\centerline{\hbox{
\psfig{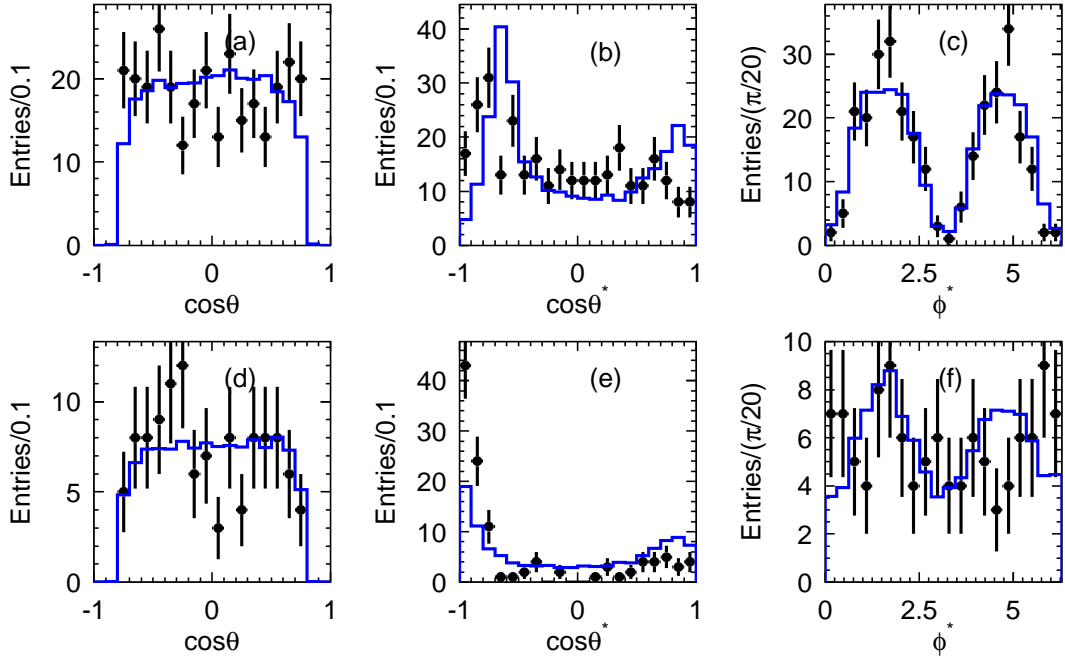}}} \caption{Angular
distributions of selected $\pspto \ppb \piz$ (top) and $\pspto
\ppb \eta$ (bottom) events. The dots with error bars are data and
the histograms are MC simulation (normalized to the same number of
events). The first through third columns are $\cos\theta$,
$\cos\theta^*$ and $\phi^*$ distributions.} \label{angdtmc}
\end{figure*}

Fig.~\ref{angdtmc} shows the angular distributions for our
selected data samples as well as the mixed MC samples, where
$\theta$ is the polar angle of the proton measured in the $\psip$
rest frame, $\theta^*$ and $\phi^*$ are the polar and azimuthal
angles of the anti-proton in the rest frame of $\overline{p}\pi$
(or $\overline{p}\eta$). The simulations are similar to data,
although not perfect. Using the same analysis as used for data on
the two mixed MC samples, with proper fractions of background
added to the $\g\g$ mass distributions, yields efficiencies of
$(14.04\pm 0.14)\%$ for $\pspto \ppb \piz$ and $(14.00\pm 0.20)\%$
for $\pspto \ppb \eta$.

\section{Systematic Errors}

The systematic errors, whenever possible, are evaluated with pure
data samples that are compared with the MC simulations.
Table~\ref{syst} lists the systematic errors from all sources.
Adding all these errors in quadrature, the total percentage errors
are 11.2\% and 11.6\% for $\piz\ppb$ and $\eta\ppb$ respectively.
The detailed analyses are described in the following.

\begin{table}[htbp]
\caption{Summary of systematic errors. Numbers common to the
two channels are only listed once.}
\begin{center}
\begin{tabular}{c|cc}\hline\hline
 Source         & ~~~~~$\piz\ppb$ (\%)~~~~~ & ~~~~~$\eta\ppb$ (\%)~~~~~\\\hline
 MC statistics  &  1.0       &  1.4  \\
 Photon ID      & \multicolumn{2}{c}{1.1}\\
 Photon efficiency    & \multicolumn{2}{c}{4}  \\
 $\piz(\eta)$ reconstruction &\multicolumn{2}{c}{2.0}   \\
 Tracking and particle ID
                &  2.6       &  2.8 \\
 Fit to mass spectrum    &  4.5       &  5.3 \\
 Decay dynamics    &\multicolumn{2}{c}{5.9} \\
 Kinematic fit         & \multicolumn{2}{c}{5}    \\
 Number of $\psip$ & \multicolumn{2}{c}{4}\\
 Trigger efficiency &\multicolumn{2}{c}{0.5}\\
 $\BR[\piz(\eta) \ra \g\g]$
                &  0.0       &  0.7 \\\hline
 Total Systematic error
                & 11.2       & 11.6 \\\hline\hline
\end{tabular}
\end{center}
\label{syst}
\end{table}

\subsection{Photon ID}

\begin{figure*}[htbp]
\centerline{\hbox{
\psfig{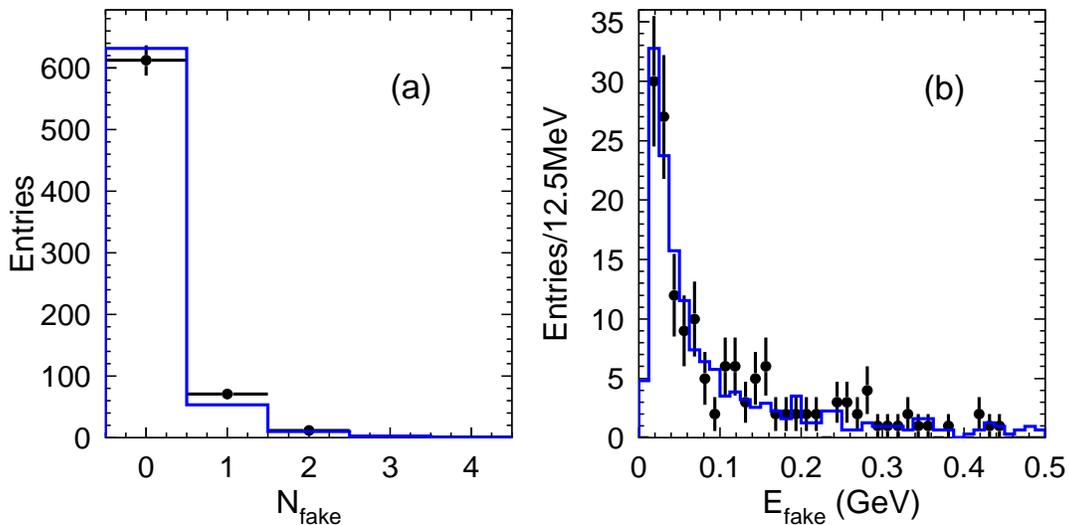}}}
\caption{Fake photon multiplicity distributions (left) and fake
photon energy spectra (right) for $\psipto \ppb$ data (dots with
error bars) and Monte Carlo simulation (histogram).} \label{nfake}
\end{figure*}

The fake photon multiplicity distributions and energy spectra for
both data and Monte Carlo simulation for $\psipto \ppb$ are shown
in Fig.~\ref{nfake}. For this decay channel, the Monte Carlo
simulates slightly less fake photons than data, while it simulates
the energy spectra reasonably well.

Using a toy Monte Carlo simulation, we found that for $(97.10\pm
0.32)\%$ of the cases, the energies of both real photons are
greater than those of the fake ones from data, while for Monte
Carlo simulation, the corresponding fraction is $(97.78\pm
0.15)\%$. A factor of $(0.993\pm0.004)$ is found between data and
Monte Carlo. We do not apply a correction to the MC efficiency;
instead, we take 1.1\% as the systematic error of photon
identification (ID).

\subsection{Photon detection efficiency}

The simulation of the photon detection efficiency is studied using
$\jpsito \pp\piz$ events with one photon missing in the kinematic
fit and examining the detector response in the missing photon
direction~\cite{lism}. The Monte Carlo simulates the detection
efficiency of data within 2\% for each photon in the full energy
range. Since we have two photons, 4\% is taken as the systematic
error of the photon detection efficiency.

\subsection{\boldmath $\piz$ and $\eta$ reconstruction}

\begin{figure}[htbp]
\centerline{\hbox{ \psfig{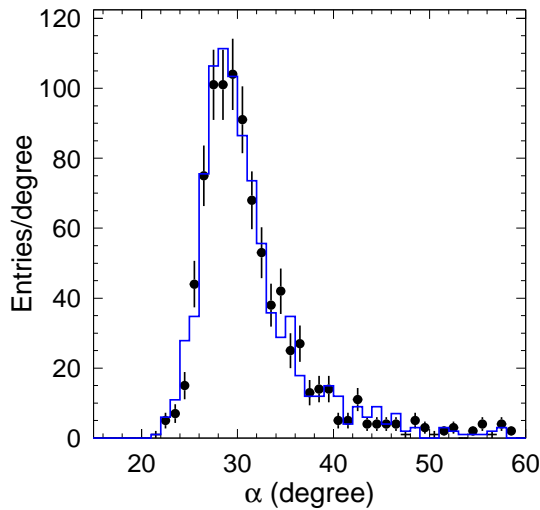}}}
\caption{Comparison of the $\g\g$ opening angle ($\alpha$)
distributions for $\piz$ from $\jpsito \pip\pim \piz$ with
momentum between 0.5~GeV/$c$ and 0.6~GeV/$c$. Dots with error bars
are data, and the histogram is Monte Carlo simulation. The
distributions are normalized to the number of events with
$\alpha>28^{\circ}$.} \label{aldis}
\end{figure}

The $\piz$ reconstruction efficiency is studied by comparing the
opening angle between the two photons between data and MC
simulation in different $\piz$ momentum ranges using $\jpsito \ppb
\piz$ and $\jpsito \pip\pim\piz$ samples selected from BES $\jpsi$
data. Fig.~\ref{aldis} shows the comparison for $\piz$ momentum
between 0.5~GeV/$c$ and 0.6~GeV/$c$, the agreement at small
opening angle shows the simulation of the $\piz$ reconstruction
efficiency is good. By reweighting the difference between data and
MC simulation in all momentum bins with the $\piz$ momentum
spectrum in $\pspto \ppb\piz$, the overall correction factor to
the MC simulation is determined to be ($98.8\pm0.8$)\%, and 2.0\%
is then taken as the systematic error due to the $\piz$
reconstruction.

The angle between the two photons emitted from the $\eta$ is
generally much greater than that from the $\piz$. As a
conservative estimation, the uncertainty for $\eta$ reconstruction
is also taken to be 2.0\%.

\subsection{MDC tracking and particle ID efficiency}

The efficiencies for protons and antiprotons that enter the
detector being reconstructed and identified are measured using
samples of $\jpsito \pip\pim\ppb$ and $\psi^\prime\rightarrow
\ppjpsi,J/\psi\rightarrow p \overline{p}$ events, which are
selected using kinematic fit and particle ID for three tracks,
allowing one proton or antiproton at a time to be missing in the
fit. The efficiency is determined by how often a proton (or
antiproton) is found in the direction of the missing track, it
varies from 80\% to around 95\% with increasing momentum, and the
MC simulates data rather well, except for proton or antiproton
momenta less than 0.5~GeV/$c$, where the nuclear interaction cross
section of particles with the detector material is very large.

The net difference between data and MC simulation is found to be
$1.001\pm 0.025$ for $\psipto \ppb \piz$, and $1.010\pm 0.018$ for
$\psipto \ppb \eta$. The errors together with the differences from
unity will be considered as systematic errors, that is, 2.6\% and
2.8\% for $\psipto \ppb \piz$ and $\psipto \ppb \eta$,
respectively.

\subsection{Fit range and background shape}

The background shape in fitting the $\g\g$ mass distributions is
changed from a second order polynomial to a first order one, and
the fit range is changed, to determine the uncertainties due to
the fitting for the $\ppb\piz$ and $\ppb\eta$ channels. Different
ways for choosing the $\piz$ momentum bins or fitting the $\piz$
signal without binning yields differences in the branching
fraction less than 3\% for the $\ppb\piz$ channel. Adding all
these in quadrature, 4.5\% and 5.3\% are taken as the systematic
error due to the fit.

\subsection{Decay dynamics}

Table~\ref{eff} shows efficiencies determined with different MC
samples; different decay dynamics result in different
efficiencies. While the mixed Monte Carlo samples with $N^*$ and
possible $\ppb$ intermediate states are used in the determination
of final selection efficiencies, the differences between the mixed
samples and the phase space generator are taken as systematic
errors due to the lack of the precise knowledge of the decay
dynamics to be used in the MC generator. The differences are found
to be 2.1\% for $\pspto \ppb \piz$ and 5.9\% for $\pspto \ppb
\eta$. The larger difference (5.9\%) will be taken as the
systematic error for both due to the uncertainty of the generator
for both channels. It should be noted that the differences between
these two MC samples in the angular distributions are large
compared with those observed in Fig.~\ref{angdtmc}; thus the
errors quoted cover the differences in the angular distribution
simulation also.

\begin{table}[htbp]
\caption{Efficiencies determined with different MC samples.}
\begin{center}
\begin{tabular}{lcc}\hline\hline
          Channel &   ~~~~~$\ppb \piz$ (\%)~~~~~
                         & ~~~~~$\ppb \eta$ (\%)~~~~~\\\hline
Only $N^*(1535)$  &   13.04            & 12.71 \\
Only $\RR$        &   18.43            & 18.40 \\
Mixed sample      &   $14.04\pm 0.14$  & $14.00\pm 0.20$ \\
Phase space       &   14.37            & 14.83 \\\hline\hline
\end{tabular}
\end{center}
\label{eff}
\end{table}

\subsection{Other systematic errors}

The uncertainty due to the kinematic fit is extensively studied
using many channels which can be selected cleanly without using a
kinematic fit~\cite{gpp,wangz,fangss,vt}. It is found that the MC
simulates the kinematic fit efficiency at the 5\% level for almost
all the channels tested. We take 5\% as the systematic error due
to the kinematic fit.

The results reported here are based on a data sample corresponding
to a total number of $\psip$ decays, $N_{\psp}$, of $(14.0 \pm
0.6) \times 10^6$, as determined from inclusive hadronic
events~\cite{pspscan}.  The uncertainty of the number of $\psip$
events, 4\%, is determined from the uncertainty in selecting the
inclusive hadrons.

The trigger efficiency is around 100\% with an uncertainty of
0.5\%, as estimated from Bhabha and $e^+ e^- \ra \mu^+ \mu^-$
events. The systematic errors on the branching fractions used are
obtained from the Particle Data Group (PDG)~\cite{pdg} tables
directly.

\section{Conclusion and Discussion}

The branching fractions of $\pspto \ppb \piz$ and $\pspto \ppb
\eta$ are calculated using
\begin{eqnarray*}
   \BR(\pspto \ppb \piz)&=&\frac{n^{obs}_{\piz}/\varepsilon}
{N_{\psp}\cdot \BR(\piz \ra \g\g)},\\
   \BR(\pspto \ppb \eta)&=&\frac{n^{obs}_{\eta}/\varepsilon}
{N_{\psp}\cdot \BR(\eta \ra \g\g)}.
\end{eqnarray*}

Using numbers listed in Table.~\ref{br}, we obtain
\begin{eqnarray*}
   \BR(\pspto \ppb \piz)& =& (13.2\pm 1.0\pm 1.5)\times 10^{-5},\\
   \BR(\pspto \ppb \eta)& =& (5.8\pm 1.1\pm 0.7)\times  10^{-5},
\end{eqnarray*}
where the first errors are statistical and the second are
systematic.  The measured $\pspto \ppb \piz$ branching fraction
agrees with Mark-II within errors~\cite{markii}.

\begin{table}[htbp]
\caption{Numbers used in the branching fraction calculation and
final results.}
\begin{center}
\begin{tabular}{c|cc}\hline\hline
quantity & $\ppb\piz$ & $\ppb\eta$\\\hline
$n^{obs}$         & $ 256  \pm 18$  & $44.8^{+8.7}_{-8.4}$\\
$\eff$ (\%)       & $ 14.04\pm0.14$ & $14.00\pm0.20$ \\
$N_{\psip}$($10^6$)   & \multicolumn{2}{c}{$14.0 \pm 0.6$} \\
$\BR(\piz(\eta) \ra \g\g$)(\%) & $98.80\pm0.03$ & $39.43\pm0.26$\\\hline
$\BR(\pspto \ppb\piz(\eta))$ ($10^{-5}$) & ~~~~$13.2\pm 1.0\pm 1.5$~~~~
                                & ~~~~$5.8\pm 1.1\pm 0.7$~~~~  \\\hline
\end{tabular}
\end{center}
\label{br}
\end{table}

Comparing the branching fractions with those of $\jpsi$ decays
from the PDG~\cite{pdg}, we find that for $\psp$ decays,
$\BR(\ppb\piz)>\BR(\ppb\eta)$, while for $\jpsi$ decays,
$\BR(\ppb\piz)<\BR(\ppb\eta)$, and
\begin{eqnarray*}
Q_{\ppb \piz}&=&\frac{\BR(\psipto\ppb \piz)}{\BR(\jpsito \ppb \piz)}=
\frac{(13.2\pm1.0\pm1.5)\times10^{-5}}{(1.09\pm0.09)\times10^{-3}}
= (12.1\pm 1.9) \%, \\
Q_{\ppb \eta}&=&\frac{\BR(\psipto\ppb \eta)}{\BR(\jpsito \ppb \eta)}=
\frac{(5.8\pm1.1\pm0.7)\times10^{-5}}{(2.09\pm0.18)\times10^{-3}}
= (2.8\pm0.7) \%.
\end{eqnarray*}
While $Q_{\ppb \piz}$ agrees well with the so-called 12\% rule,
$Q_{\ppb \eta}$ seems to be suppressed significantly.

\begin{figure*}[htb]
\centerline{\hbox{
\psfig{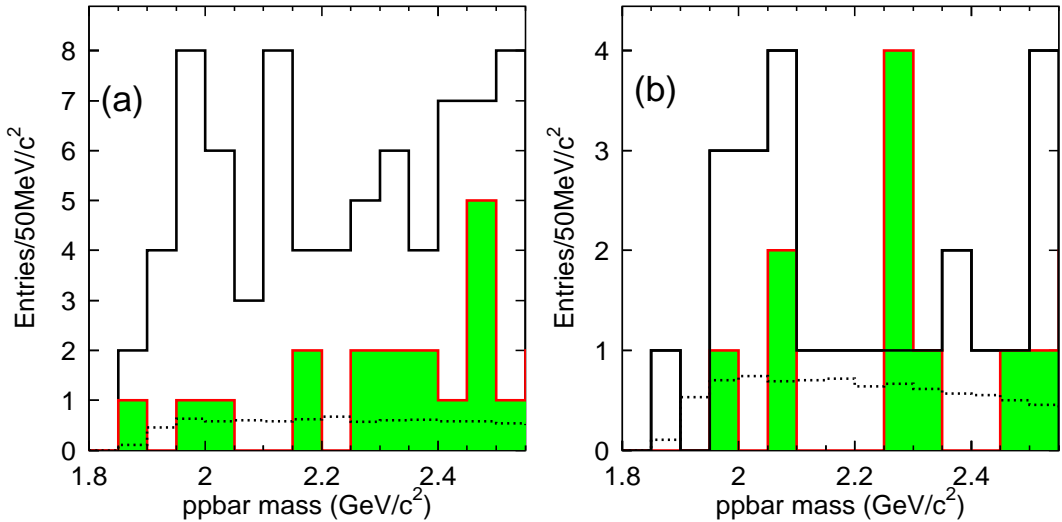}}} \caption{$\ppb$
invariant mass distributions of selected (a) $\ppb\piz$  and (b)
$\ppb\eta$  events. The blank histograms are selected signal
events, and the shaded histograms are events from $\piz$ or $\eta$
mass sidebands. The dashed histograms are predictions of phase
space with S-wave $\ppb$ (not normalized).} \label{mppb}
\end{figure*}

Fig.~\ref{mppb} shows the $\ppb$ invariant mass distributions of
the selected $\ppb\piz$ and $\ppb\eta$ events shown in
Fig.~\ref{dalitz}, together with the expected background estimated
from $\piz$ or $\eta$ mass sidebands (0.075-0.100 and
0.170-0.195~GeV/$c^2$ for $\piz$ and 0.49-0.51 and
0.59-0.61~GeV/$c^2$ for $\eta$). There are indications of some
enhancement around 2~GeV/$c^2$ in both channels. Fitting the
enhancement with an S-wave Breit-Wigner and a linear background,
with a mass dependent efficiency correction, yields a mass around
2.00~GeV/$c^2$ in the $\ppb\piz$ mode and 2.06~GeV/$c^2$ in
$\ppb\eta$, with the width in both channels around
30-80~MeV/$c^2$, and significance around 2.7$\sigma$. Fitting with
a P-wave Breit-Wigner results in slightly lower masses and similar
significance. The nature of the enhancements is not clear, and the
statistics are too low to allow a detailed study. The enhancements
in the two channels cannot be the same since they have different
isospin.

Fig.~\ref{res2} shows projections of Dalitz plots in $p\pi$ (or
$p\eta$) invariant mass after removing $\pspto \jpsi + X$
backgrounds and the possible $\ppb$ mass threshold enhancements.
There is a faint accumulation of events in the $p\pi$ invariant
mass spectrum at around 2065~MeV/$c^2$, but it is not
statistically significant. The enhancement between 1.4 and
1.7~GeV/$c^2$ may come from $N^*(1440)$, $N^*(1520)$, $N^*(1535)$,
etc. We do not attempt a partial wave analysis due to the limited
statistics. There is a clear enhancement with $p\eta$ mass at
$(1549\pm 13)$~MeV/$c^2$, which is possibly the $N^*(1535)$.

\begin{figure*}[htb]
\centerline{\hbox{ \psfig{file=res_ycz_fig8.epsi,width=14.0cm}}}
\caption{Projections of Dalitz plots in $p\pi(p\eta)$ invariant
mass after removing $\pspto \eta\jpsi$ and the possible $\RR$. (a)
and (b) are $m_{p\piz}$ and $m_{\overline{p}\piz}$ in $\pspto \ppb
\piz$; (c) is the sum of (a) and (b); (d) and (e) are $m_{p\eta}$
and $m_{\overline{p}\eta}$ in $\pspto \ppb \eta$; (f) is the sum
of (d) and (e).} \label{res2}
\end{figure*}

\section{Summary}

$\ppb\piz$ and $\ppb\eta$ signals are observed in $\psip$ decays,
and the corresponding branching fractions are determined. For
$\psipto\ppb\piz$, the errors are much smaller than those of the
previous measurement by Mark-II~\cite{markii}, and for
$\psipto\ppb\eta$, it is the first observation. There is no clear
$N^*(2065)$ peak in the $\ppb\piz$ mode, but there is some weak
evidence for $\ppb$ threshold enhancements in both channels.

\acknowledgments

The BES collaboration thanks the staff of BEPC for their hard
efforts. This work is supported in part by the National Natural
Science Foundation of China under contracts Nos. 10491300,
10225524, 10225525, the Chinese Academy of Sciences under contract
No. KJ 95T-03, the 100 Talents Program of CAS under Contract Nos.
U-11, U-24, U-25, and the Knowledge Innovation Project of CAS
under Contract Nos. U-602, U-34 (IHEP); by the National Natural
Science Foundation of China under Contract No. 10175060 (USTC),
and No. 10225522 (Tsinghua University); and by the Department of
Energy under Contract No. DE-FG02-04ER41291 (University of
Hawaii).

\end{document}